\pdfoutput=1
\documentclass[a4paper,10pt, twocolumn]{article}

\usepackage{amsmath,amssymb,amsfonts}
\usepackage{algorithmic}
\usepackage{graphicx}
\usepackage{textcomp}
\usepackage{xcolor}
\usepackage{booktabs}
\usepackage{multirow}
\usepackage{array}
\usepackage{subfig}
\usepackage{fixltx2e}

\newcommand{\figH}{2.5in}
\newcommand{\figSH}{1.0in}
\newcommand{\sfpt}{\emph{$SmartFPTuner$}}
\newcommand{\sfptplus}{\emph{$SmartFPTuner^+$}}

\date{}

\begin{document}

\title{Combining Learning and Optimization for Transprecision Computing}


\author{Andrea Borghesi, Giuseppe Tagliavini, Michele Lombardi, Luca Benini,
Michela Milano \\DISI/DEI University of Bologna}

\maketitle              

\begin{abstract}
The growing demands of the worldwide IT infrastructure stress the need for
reduced power consumption, which is addressed in so-called transprecision
computing by improving energy efficiency at the expense of precision. For
example, reducing the number of bits for some floating-point operations 
leads to higher efficiency, but also to a non-linear decrease of the
computation accuracy. Depending on the application, small errors can be
tolerated, thus allowing to fine-tune the precision of the computation.  
Finding the optimal precision for all variables in respect of an error 
bound is a complex task, which is tackled in the literature via heuristics. In
this paper, we report on a first attempt to address the problem by combining a
Mathematical Programming (MP) model and a Machine Learning (ML) model, following the
Empirical Model Learning methodology. The ML model learns the relation between
variables precision and the output error; this information is then embedded in
the MP focused on minimizing the number of bits. An additional refinement
phase is then added to improve the quality of the solution. The experimental
results demonstrate an average speedup of 6.5\% and a 3\% increase in solution
quality compared to the state-of-the-art. In addition, experiments on a
hardware platform capable of mixed-precision arithmetic (PULPissimo) show the
benefits of the proposed approach, with energy savings of around 40\% compared
to fixed-precision.
\end{abstract}

\section{Introduction} \label{sec:intro}

The energy consumption of computing systems keeps growing and, consequently,
considerable research efforts aim at finding energy-efficient solutions. A wide
class of techniques belongs to the \emph{approximate
computing}\cite{xu2016approximate} field, which has the goal of decreasing the
energy associated with computation in exchange for a reduction in the quality of
the computation results. In this area, a wide range of techniques have been
designed, from specialized HW solutions to SW-based
methods\cite{mittal2016survey}.  In recent years, a new paradigm called
\emph{transprecision computing}
emerged\cite{malossi2018transprecision,oprecompProject}, where errors are not
merely ``tolerated'' as byproducts, but rather SW and HW solutions are designed
to provide the desired computation quality. Floating-point (FP) operations are
a common target for transprecision techniques, as their execution and related
data transfers represent a large share of the total energy consumption for many
applications involving a wide numerical
range\cite{klavik2014changing,chiang2017rigorous}. For instance, Tagliavini et
al. developed \emph{FlexFloat} \cite{tagliavini2018transprecision}, an
open-source SW library that allows to specify the number of bits used for the
mantissa and the exponent of an FP variable: using a smaller number of bits
decreases the precision, thus saving energy. 

With the possibility to fine-tune the precision of application variables comes
the challenge of finding the best setup. This can be framed as an optimization
problem, solvable by paradigms such as Mathematical Programming (MP). The idea
is to search for the minimal number of bits that can be assigned to each
variable without incurring in a computation error larger than a target. This
method requires to analytically express the non-linear relation between
precision and the computation error, not a trivial task
\cite{moscato2017automatic}. Static analysis of the effect of variables
precision is burdensome, and most current approaches have severe
limitations\cite{Darulova:2017:TCR:3062396.3014426,chiang2017rigorous}.  A
possible solution is to \emph{learn}, rather than directly express, this
relation via ML models. We could then embed this knowledge in the optimization
model and solve it. This notion is the core idea of \emph{Empirical Model
Learning} (EML)\cite{lombardi2017empirical}, a technique to enable combinatorial
optimization over complex real-world systems. 

In this paper, we propose a novel optimization method to find optimal variable
precision in a transprecision computing setting, based on the EML methodology.
The main contributions of this work are: 
\begin{enumerate} 
    \item A novel approach, called \sfpt, that combines ML models (predicting
        the error associated with variable precision) and an MP optimization
        model (finding the optimal precision under a constraint on the error) --
        this method provides a 55\% reduction in solution time w.r.t.
        state-of-the-art (SoA) tools;
    \item An extended model, called \sfptplus, that trades off quickness for 
        quality and merges the optimization approach with the SoA algorithm, 
        obtaining a 6.5\% speedup over the SoA and a 3\% increase in
        solution quality.
\end{enumerate}

\sfpt~enables a significant improvement in execution time that allows
integrating this approach into compilation toolchains, but in some cases it
produces solutions of lower quality and with marginal energy benefits;
\sfptplus~bridges this gap, always providing good execution time and
high-quality solutions at the same time.
Further experiments on PULPissimo, an ultra-low-power platform provided with a
mixed-precision HW FP unit, show additional energy savings around 40\%.


The rest of the paper is organized as follows. Section~\ref{sec:related}
discusses the related work in FP precision tuning. Section~\ref{sec:approach}
introduces the proposed approach. Sections~\ref{sec:exp_deploy} and
\ref{sec:fptune_exp} show experimental results on precision tuning and energy
efficiency, respectively. Finally, Section~\ref{sec:conclusion} provides 
conclusion and future directions.

\section{Related Work} \label{sec:related}
Several works in recent years proposed specific algorithms for FP
variable precision tuning
\cite{graillat2016promise,rubio2016floating,guo2018exploiting}.
The current SoA is the parallel algorithm called \emph{fpPrecisionTuning}, 
proposed by Ho et al.\cite{ho2017efficient}; it is an  automated tool that
fine-tunes the number of bits to be assigned to FP variables while
respecting the constraint on the desired maximum error (for brevity, we
refer to this algorithm as \emph{FPTuning} in the rest of the paper). The
algorithm searches for the best solution by running the application
to be tuned with different precision levels; a binary search algorithm 
explores the precision ranges.

Many works have tried to analyze the error introduced by tuning FP
variables\cite{rubio2016floating,moscato2017precisa,chiang2017rigorous}. While
promising, these approaches suffer from some limitations: they mostly work at
the single-expression level and cannot handle whole benchmarks; those dealing
with entire programs (e.g., \cite{rubio2016floating}) are orders of magnitudes
slower than methods such \emph{FPTuning}; they consider only very few precision
levels (e.g., single or double precision).

Finding optimal parameter values for a given algorithm is a well-known area of
research. For example, Hutter et al. propose a Sequential Model-based
optimization for general Algorithm Configuration (SMAC,
\cite{hutter2011sequential}), an automated procedure for algorithm
configuration that explores the space of parameter settings. The approach
relies on building regression models that describe the relationship between the
target algorithm performance and the configuration. Our problem can be cast in
the SMAC scheme if we treat the precision of the variables as the algorithm
configurations to be explored and the desired target error as a bound on the
algorithm performance. We applied SMAC to our problem, but the preliminary 
attempts were computationally expensive, and the resulting quality lower than 
problem-specific techniques.
Costa et al. developed RBFOpt \cite{costa2018rbfopt}, an open-source library
for optimization with black-box functions. The method iteratively refines a
kernel-based surrogate model of the target function, which is used to guide the
search. Our task can be seen as a black-box optimization problem by considering
the precision values as the decision variables and the error as the black-box
function. 

Empirical Model Learning is a relatively new research area, with many
potential applications \cite{BartoliniLMB12}. We are
particularly interested in two specific works, namely: I) Lombardi et al.
\cite{BartoliniLMB11}, which shows how to embed a neural-network-based model in
a combinatorial problem, and II) Bonfietti et al. \cite{BonfiettiLM15}, 
integrating Decision Trees (DT) and Random Forest (RF) models within an MP
problem. In our approach, we use their contributions to embed the ML models for
predicting the error associated with the variable precision. 

\section{Proposed Approach} \label{sec:approach}

\subsection{Problem Description}
\label{sec:approach_problem_description}

We consider numerical benchmarks where multiple FP variables take part in 
the computation of the result for a given input set, which includes a 
structured set of FP values (typically a vector or a matrix). The number of variables
with controllable precision in a benchmark $B$ is $n_{var}^B$; these variables 
are the union of the original variables of the program and the additional variables 
inserted for handling the intermediate results. For example, if in the original 
program we have an instruction $V_1 = V_2 + V_3$ involving three FP variables, 
the set $B$ contains \emph{four} FP variables, three corresponding to the 
$V_i$ variables plus the temporary one added to match the precision of the sum 
before the assignment (i.e., the precision at which the operation is performed). 
Adopting this approach, each variable is free to contribute to multiple expressions 
with different precision; practically, HW arithmetic units require operands 
of the same type, and this requirement can be satisfied with the additional variable.

Our problem consists of assigning a precision to each FP variable while
respecting a constraint on the computation error. Assigning a precision means
deciding the number of bits for the mantissa; the exponent dictates the
extension of dynamic range and is set according to the actual types
available on the target HW platform. In the rest of the paper, we refer to the 
reduction in output quality due to the adjusted precision
(reduction w.r.t. the output obtained with maximum precision) as
\emph{error}. If $O$ indicates the result computed with the fine-tuned precision and
$O^M$ the one obtained with maximum precision, we compute the error $E$ as: 
$E = \max_{i} \frac{(o_i - o_i^M)^2}{(o_i^M)^2}$
This error metric has been adopted by the current SoA algorithm for precision 
tuning \cite{ho2017efficient}, and it is one of a broad set of metrics proposed for
transprecision computing\cite{malossi2018transprecision,oprecompProject}.

In this approach, we focus on the single input set case: we assume a fixed input
set to be fed to the benchmark, and we look for the best solution given that
precise input set. Consequently, the optimal solution for an input set may not
respect the error constraint for other ones. This requirement is not an issue
for the comparison with the SoA as it makes the same assumption; our future work
aims at overcoming this limitation.

We selected a subset of the applications studied in the context of
transprecision computing\cite{oprecompProject}, chosen because
they represent distinct problems and capture different patterns of computation.
At this stage, we do not consider whole applications (i.e., training a deep
neural net) but we focus on micro-benchmarks that are part of larger
applications (i.e., convolution operations, matrix multiplications, etc.).  In
particular, we chose the following benchmarks: 
\begin{itemize} 
    \item \emph{FWT}, Fast Walsh Transform for real vectors, from
    the domain of advanced linear algebra ($n_{var}^{FWT} = 2$); 
    \item \emph{saxpy}, a generalized vector addition with the form $y_i = a \times
    x_i + y_i$, basic linear algebra ($n_{var}^{saxpy} = 3$); 
    \item \emph{convolution}, convolution of a matrix with a $11 \times 11$
        kernel, ML ($n_{var}^{conv} = 4$); 
    \item \emph{dwt}, Discrete wavelet
    transform, from signal processing ($n_{var}^{dwt} = 7$); 
    \item \emph{correlation},
    compute correlation matrix of input, data mining ($n_{var}^{corr} = 7$).
    \item \emph{BScholes}, estimates the price for a set of options applying
    Black-Scholes equation, from computational finance ($n_{var}^{BScholes} =
    15$); 
    \item \emph{Jacobi}, Jacobi method to track the evolution of a 2D
    heat grid, from scientific computing ($n_{var}^{Jacobi} = 25$).
\end{itemize}

We stress out that this is a complex problem, especially the relationship
between variable precision and error. First, the error measure is very
susceptible to differences between output at maximum precision and output at
reduced precision, due to its maximization component. Secondly, the
precision-error space is non-smooth, non-linear, non-monotonic, and with many
peaks (local optima). In practice, increasing the precision of all variables
does not guarantee to reduce the error. This effect is due to multiple factors, such as
the impact of rounding operations and the effects of numerical stability on the
control flow\cite{Darulova:2017:TCR:3062396.3014426}.  
For instance, suppose to increase only the precision of a
variable involved in the condition of an \emph{if} statement with a constant FP value.  
Since the modification does not consider this dependence, a rounding of the 
variable (when its value is near the constant) can trigger different code 
branches and produce unexpected results on the output.

\subsection{Approach Description}
\label{sec:approach_description}

We propose an optimization model based on three components: 1) an MP model, 2)
an ML model to predict the error associated with a precision configuration, and 3)
an ML model to classify configurations in two macro-classes based on the
associated error (i.e., \emph{small} or \emph{large}). The two ML models are
embedded in the MP model and represent the knowledge about the relationship between
variables precision and output error. 

\begin{figure}[hbt]
	\centering
    \includegraphics[width=0.45\textwidth,height=\figH]{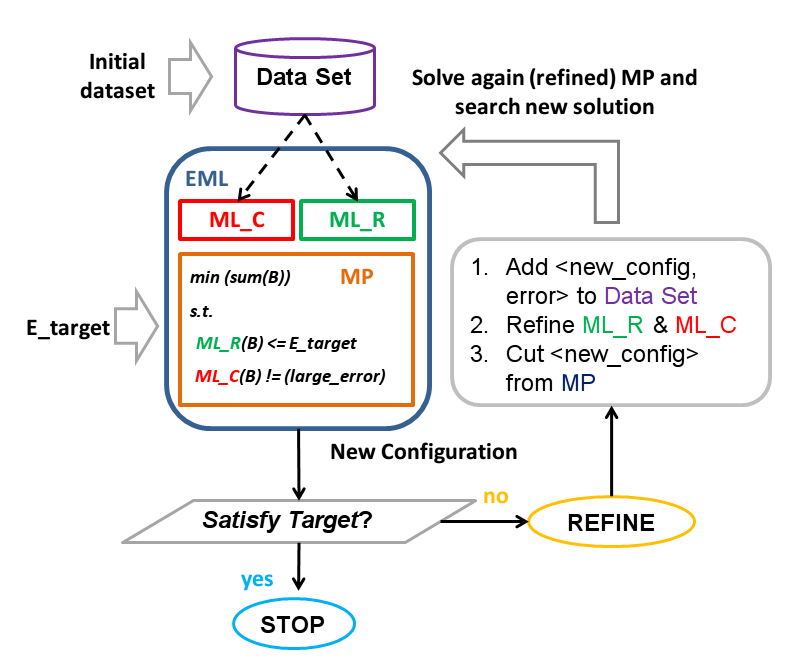}
    \caption{Scheme of the Approach}
	\label{fig:approach_scheme}
\end{figure}

The MP model finds the optimal bit configuration according
to the prediction of the two ML models; to assess the quality of the
configuration, we execute the benchmark with the corresponding
precision. For this purpose, we employed \emph{FlexFloat}
\cite{tagliavini2018transprecision}, that allows us to run a benchmark
specifying the precision of each FP variable. The task of
the ML models is very hard since their goal is to learn a complex
function. Hence, the solution found by the MP can be unfeasible; namely, it does
not respect the constraint on the target error, due to the gap between estimated
and actual error. To fix this problem, we introduce a \emph{refinement loop}: we
test the MP solution by running the benchmark with the specified precision; if
the solution is unfeasible, we search for a new one. The wrong one (the
configuration plus its actual error) is added to the training set of the ML
models, which are then retrained, and cut from the pool of possible solutions of
the MP (via a set of constraints). A new MP model is then built based on the
refined ML models, and a new search begins; this loop goes on as long as a
feasible solution is found. The overall approach is depicted in
Figure~\ref{fig:approach_scheme}.

\subsubsection{ML Models}
\label{sec:approach_regr_class}

As a first step, we created a collection of data sets containing examples of our
benchmarks run at different precision, with the corresponding error values.
The configurations used in each data set were obtained via Latin Hypercube
Sampling\cite{stein1987large}, to explore the design space efficiently. 

\begin{figure}[hbt]
	\centering
    \includegraphics[width=0.45\textwidth,height=\figH]{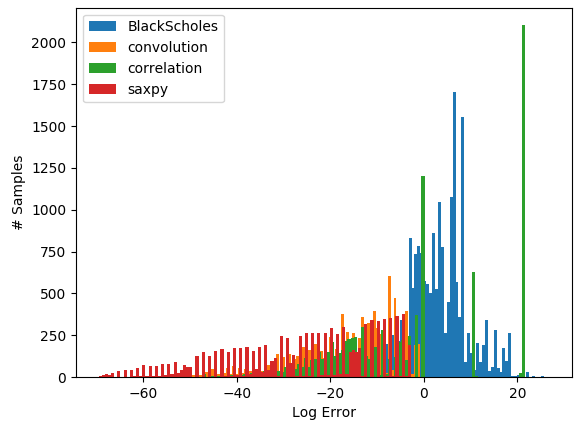}
    \caption{Prediction errors distribution in logarithmic scale}
	\label{fig:multi_log_error_100hist_noDWT}
\end{figure}

The majority of configurations lead to small errors, from $10^{-1}$ to
$10^{-30}$, as the output with fine-tuned variables does not differ drastically
from the target one. However, in a minority of cases lowering the precision of
critical variables generates errors higher than $100$. Formally, the errors 
roughly follow a long-tailed distribution: this
can be observed by plotting the histogram of the logarithmic error $\log(E)$,
as done in Figure~\ref{fig:multi_log_error_100hist_noDWT} for four of our
benchmarks. Benchmarks with fewer variables (such as \emph{saxpy} and
\emph{conv}) have a regular trend, with logarithmic errors always smaller
than $0$. When the number of variables increases, for instance with the \emph{corr}
benchmark (green bars), the majority of errors still have a logarithm
smaller than $0$; however, we can notice two spikes around $10$ and after $20$. The
situation gets even more complicated with \emph{BScholes} (blue bars); in 
this case a vast number of configurations correspond to significant errors.
This kind of output distribution makes it very difficult for a single model
trained in a classical fashion (e.g., for minimum Mean Squared Error) to provide
consistently good predictions.

Overly large error values are usually due to numerical issues arising during
computation (e.g., overflow, underflow, division by zero, or not-a-number
exceptions). This intuitively means that the large-error configurations are
likely to follow a distinct pattern w.r.t. the configurations having a 
lower error value. \emph{Accordingly, it makes sense to split the prediction
task into two specialized models}: a classifier $ML_C$ to screen large errors,
plus a regressor $ML_R$ to evaluate those configurations not ruled-out
by the classifier. The $ML_C$ needs to make a distinction between
\emph{normal error} and \emph{large error} configurations. We trained this model
by labeling each error in our data set with a class field $c$, equal to $1$ if
the error of the example is greater than a threshold ($0.9$ in our experiments),
and equal to $0$ otherwise. Configurations classified with class $c = 1$ can be
discarded by the training set of the regression model.

$ML_R$ has the task of predicting the output
error for an assignment of precision values. We
quickly noticed that any ML model we tried struggled with discerning between
small and relatively close errors (i.e., $10^{-20}$ and $10^{-15}$); therefore, we
opted to predict the negative of the logarithm of the error, thus magnifying
the relative differences and dramatically improving the ML model accuracy.
$ML_C$ and $ML_R$ will be used in the MP model with the aim to, respectively, avoid
large-error configurations and enforce the bound on the precision of the variables.
Together, the two models offer a more robust prediction, but still not a
perfect one. 

\subsubsection{MP Model} \label{sec:approach_MP}

The MP model assigns a precision value to each variable in the benchmark, and it
minimizes the total number of bits while respecting the upper bound on the error.
We have an integer decision variable $x_i^B, \forall i \in \{1\ldots
n_{var}^B\}$, namely for each variable of the benchmark. The
decision variables represent the number of bits assigned to the variable $x_i
\in \{nbit_{min}\ldots nbit_{max}\}$. Then we have a continuous variable $e$
that represents the error predicted by $ML_R$; as specified earlier, the
predicted error is the negative log of the actual error. Finally, we have a
variable $c \in \{0,1\}$ which stands for the output of the classifier $ML_C$.
The decision variables $x_i$ and the $e$ and $c$ variables are connected by a
set of constraints that encode the $ML_C$ and $ML_R$ models, generated via the
EML library EMLlib\footnote{https://github.com/emlopt/emllib}.

We then add the constraint that forces the solver not to choose precision
values leading to large errors, namely we require $c = 0$. 
We bound the (predicted) error to be below a given target
($E^{target}$, again expressed as log) and then we minimize the total number
of bits assigned to the variables: 
\begin{align}
    \min z & =  \sum_{i=1}^{n_{var}^B} x_i \\
    \text{s.t. } & e \geq E^{target} \\
    \text{s.t. } & c = 0
\end{align}
It is important to notice that the constraints described by Equations (1-2)
depend on EML methodology, as they encapsulate the empirical knowledge obtained
through the ML models. The actual formulation of these constraints has been
omitted due to space limitations, as embedding an ML model can require up to
hundreds of even thousands of constraints. Nevertheless, the full
implementation of the MP model is available on a public code
repository
\footnote{https://github.com/oprecomp/StaticFPTuner}.
Generally
speaking, the number of constraints added due to the embedding of ML models
inside MP optimization problems strongly depends on the number of variables in a
benchmark, ranging from $38$ in the case of \emph{FWT} to $4235$ in the case of
\emph{Jacobi} (for an intermediate benchmark such as \emph{dwt} the number of
additional constraints is equal to $513$). We refer to several works already
published\cite{BartoliniLMB12,LombardiG13,LombardiG16,BonfiettiLM15}
and the publicly available code for details on how ML models can be embedded in
MP models as a set of additional constraints.

\begin{figure}[hbt]
	\centering
    \includegraphics[width=0.2\textwidth,height=\figSH]{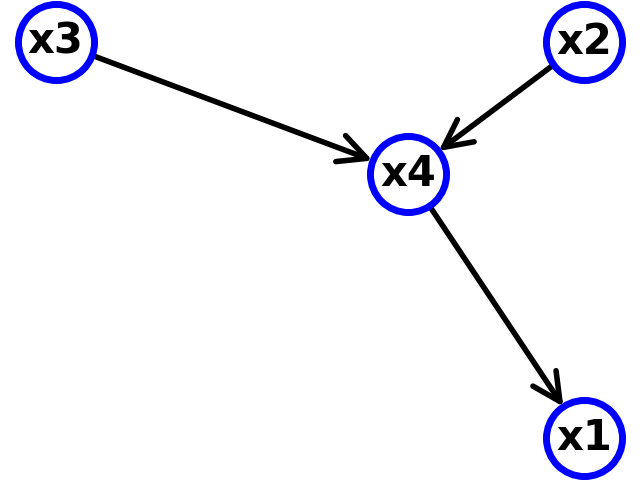}
    \caption{Example of Dependency Graph}
	\label{fig:depGraph_example}
\end{figure}

An additional set of constraints derives from the \emph{dependency graph} of the
benchmark, which specifies how the program variables are related. For instance,
consider again the expression $V_1 = V_2 + V_3$; this corresponds to four
precision levels that need to be decided $x_i, i \in [1,4]$. The first three
precision-variables $x_1$, $x_2$, and $x_3$ correspond to the precision of the
actual variables of the expression, respectively $V_1$, $V_2$, and $V_3$; the
last variable $x_4$ is a \emph{temporary} precision-variable introduced with the
\emph{FlexFloat} API to handle the (possibly) mismatching precision of the
operands $V_2$ and $V_3$ (\emph{FlexFloat} performs a cast from $x_2$ and $x_3$
to the intermediate precision $x_4$). Each variable is a node in the dependency
graph, and the relations among variables are directed edges, as depicted in
Fig.~\ref{fig:depGraph_example}; an edge entering a node means that the
precision of the source-variable is linked to the precision of the
destination-variable.

From this graph, we can extract additional constraints for the MP model; these
constraints greatly prune the search space, thus massively reducing the time
needed to find a solution. We focus on two types of relations: I) assignment (e.g.,
$x_4 \rightarrow x_1$), and II) expression-induced cast (e.g., $x_2,x_3
\rightarrow x_4$), meaning that the result of an expression involving multiple
variables has to converge to the precision associated to the additional variable
$x_4$. 

In assignment expressions, we impose that the precision of the value to be
assigned needs to be smaller or equal to the precision of the result, in our
example: $x_4 \leq x_1$. Assigning a larger number of bits to the value to be
assigned $x_4$ would be pointless since the final precision of the expression is
ultimately dependent on the precision of the result variable ($x_1$). For
relations of the second type, we instead bound the additional variable to have a
precision equal to the minimum precision of the operands involved in the
expression ($x_4 = min(x_2, x_3)$).

\section{Experimental Results: Precision Tuning} \label{sec:fptune_exp}
In this section we provide the implementation of the approach for the selected 
benchmarks, providing an evaluation of execution time and solution quality.

\subsection{ML Models} \label{sec:exp_ML_explore}
The current version of the EML library supports two types of ML models, Decision
Trees (DT) and Neural Networks (NN): We considered both these techniques in 
our exploration. The DT and NN models are implemented,
respectively, with \emph{scikit-learn} ML Python module and with \emph{Keras}
and \emph{TensorFlow}. The NNs are trained with \emph{Adam}
\cite{kingma2014adam} optimizer with standard parameters; the number of epochs
used in the training phase is 100, and the batch size is 32. 

We opted to implement $ML_R$ with a NN. After an empirical evaluation, we
realized that both NN and DT guaranteed similar prediction errors but with
different model complexities: with the NN, few simple layers were needed to reach
small errors while good DTs had to be very deep (between 40 and 50 levels).
Since the size of a DT (and its encoding) grows exponentially with depth, having
so many layers caused issues when constructing the $MP$ model; these issues are
solved by the  more straightforward structure needed by NN models. Our NN is
composed of one input layer (number of neurons equal to $n_{var}^B$), three
dense hidden layers (with size $2 \times n_{var}^B$, $2 \times n_{var}^B$, and
$n_{var}^B$), and a final output layer of size $1$; all layers employ standard
Rectified Linear Units (ReLU), except for the output layer that is linear. 

As noted before, we are not interested in having perfect error prediction
accuracy in this phase, as \sfpt~handles wrong predictions through the
refinement phase. Creating a training set needs considerable time, as it
requires the execution of multiple configurations. Hence, we use a relatively
small training set (1k examples); empirical experiments revealed that more
extensive training sets marginally increase the prediction accuracy but not
enough to justify the increase in the creation time. The average, normalized
error with this training set size and NN is around 8\%, though it varies
significantly from benchmarks with fewer variables (e.g., 4\% for \emph{saxpy})
to more complex ones (17\% for \emph{Jacobi}).

For $ML_C$, after a preliminary evaluation, we settled for using 
DTs since they provide higher accuracy than NNs even with
modest depth ($15$ in our final implementation); averaging on all benchmarks,
the $ML_C$ accuracy for DT and NN implementations are, respectively, 97\% and
82\%.

\subsubsection{Data set size and prediction error} 
\label{sec:exp_ML_dataSet_sizeImpact}

\begin{figure}[hbt]
	\centering
    \includegraphics[width=0.45\textwidth]{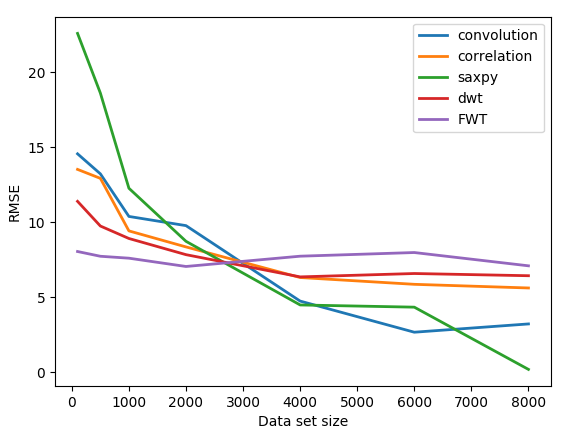}
    \caption{Data set size impact on RMSE}
	\label{fig:data_set_size_impact_pred_error}
\end{figure}

Since generating the data set used for the ML tasks has non-negligible costs
(each benchmark has to be run with many configurations), understanding the
impact of the data set size on the prediction error is crucial.
Figure~\ref{fig:data_set_size_impact_pred_error} shows the effect of the training
set size on the prediction error, measured as RMSE (one line for each
benchmark). As expected, the error decreases when the training sets contain
more examples; however, after a certain size, the gains become marginal (around
4 or 5 thousand examples).

\subsubsection{Error Classification} \label{sec:exp_ML_explore_C}

For our classifier, after an empirical evaluation we settled for using a
Decision Tree (DT): this proved to reach better accuracy w.r.t. NNs, even with modest 
depth ($20$ in our final implementation). 
Table~\ref{tab:classr_res}
compares the prediction accuracy of DT and NN classifiers (same topology as the regressor
one) for different data set sizes. The DTs models neatly
outperform NNs, strengthening our conjecture that normal errors and large
errors indeed follow different patterns. Furthermore, increasing the training set
size does not dramatically improve the performance of the classifier; smaller
training sets (around 1000 examples) can be used with good results. 

\begin{table*}
\sf\centering
\begin{tabular}{l ccccc ccccc}
 \toprule
 \multirow{3}{*}{\emph{Benchmark}} 
     & \multicolumn{5}{c}{NN} & \multicolumn{5}{c}{DT} \\
     & \multicolumn{5}{c}{Data set sizes} \qquad 
         & \multicolumn{5}{c}{Data set sizes} \\
     & 100 \quad & 500 \quad & 1k \quad & 2k \quad & 8k \qquad 
     & 100 \quad & 500 \quad & 1k \quad & 2k \quad & 8k \\ 
  \midrule
\emph{saxpy} 
     & 1.000 \quad & 1.000 \quad & 1.000 \quad & 1.000 \quad & 1.000 \qquad 
     & 1.000 \quad & 1.000 \quad & 1.000 \quad & 1.000 \quad & 1.000 \\
\emph{convolution} 
     & 1.000 \quad & 1.000 \quad & 1.000 \quad & 1.000 \quad & 1.000 \qquad
     & 1.000 \quad & 1.000 \quad & 1.000 \quad & 1.000 \quad & 1.000 \\
\emph{FWT} 
     & 0.850 \quad & 0.860 \quad & 0.660 \quad & 0.677 \quad & 0.980 \qquad 
     & 0.996 \quad & 0.997 \quad & 0.996 \quad & 0.998 \quad & 0.998 \\
\emph{correlation} 
     & 0.750 \quad & 0.790 \quad & 0.795 \quad & 0.825 \quad & 0.962 \qquad 
     & 0.996 & 0.998 \quad & 0.995 \quad & 0.991 \quad & 0.996 \\
\emph{dwt} 
     & 0.650 \quad & 0.860 \quad & 0.930 \quad & 0.618 \quad & 0.965 \qquad 
     & 0.991 \quad & 0.987 \quad & 0.989 \quad & 0.992 \quad & 0.991 \\
\emph{BScholes} & 0.700 & 0.600 & 0.675 & 0.677 & 0.827 \qquad 
           & 0.983 & 0.984 & 0.981 & 0.985 & 0.988 \\
\emph{Jacobi} & 0.750 & 0.810 & 0.800 & 0.800 & 0.836 \qquad 
           & 0.906 & 0.916 & 0.919 & 0.912 & 0.918 \\
  \midrule
\emph{Average} 
     & 0.740 \quad & 0.784 \quad & 0.772 \quad & 0.719 \quad & 0.914 \qquad 
     & 0.974 \quad & 0.976 \quad & 0.976 \quad & 0.976 \quad & 0.978 \\
  \bottomrule
\end{tabular}
\caption{$ML_C$ Accuracy Results: DT VS NN}
\label{tab:classr_res}	
\end{table*}

\subsection{MP Results} \label{sec:exp_MP_result}

We now examine the solutions found by\sfpt. All the experiments were performed
using a quad-core processor (Intel i7-5500U CPU 2.40 GHz) with 16 GB of RAM. The
MP model was solved using IBM ILOG CPLEX 12.8.0, via the Python API.  

\subsubsection{Comparison with the State-of-the-Art}
\label{sec:exp_MP_result_SoA_comp}

We compare our approach with the SoA technique for our problem, the
\emph{FPTuning} algorithm. \emph{FPTuning} proceeds by testing several precision
configurations via binary search; the algorithm is highly parallelized and leads
to solutions which are very close to the optimal one, but it has a considerable
drawback, namely it has to run the benchmark multiple times to find a feasible
solution (we can see it as a variant of a generate-and-test method). 

We can highlight two main advantages of \sfpt. First, it is more flexible
compared to a specific algorithm and more expressive, as more sophisticated
constraints can be enforced. For instance, we can constrain the precision of
the variables to assume values available on real HW implementations (typically,
the only allowed values are 16, 32, and 64 bits). Moreover, the MP can be easily
extended for architecture-specific optimization (vectorial instruction sets) and
for handling more complex objectives (e.g., minimize the number of casting
operations). Secondly, once the ML models have been trained and embedded, the MP
model can be used multiple times, relying only on the solver without the need to
perform additional benchmark runs. For example, this can be exploited to
characterize the error/precision Pareto front, whereas \emph{FPTuning} would
need to start \emph{ab initio} every time. Considering the current limitations
of \sfpt, it does not always find good solutions compared to \emph{FPTuning}; on
the contrary, the solutions of \sfpt~usually have a higher number of bits.  

Table~\ref{tab:res_cmp} provides an overview of the comparison for all
benchmarks. The values reported are computed over all error targets considered,
namely $10^{-30}$, $10^{-25}$, $10^{-20}$, $10^{-15}$, $10^{-10}$, $10^{-7}$,
$10^{-5}$, $10^{-3}$, $10^{-1}$. Each column from 2-6 corresponds to a
benchmark; the last one on the right is the average on all benchmarks. The first
row reports the difference (as a percentage) in solution quality between
\sfpt~and \emph{FPTuning}, in terms of the number of bits in the solution; a
minus sign indicates that our method outperforms \emph{FPTuning}.
The time required to find a solution by \sfpt~includes two components: I) the time
needed to create the data set to train the ML models and II) the actual solution
time, that is the time required to train and integrate the ML models, solve the
MP model and eventually repeat the process in case the solution found does not
respect the bound on the error. The second row in Table~\ref{tab:res_cmp}
reports the time difference between \sfpt~and \emph{FPTuning}, computed
excluding the time needed to create the training sets; including it would not be
fair, as after the data set is created it can be reused multiple times and
different error targets (it can be used to train different ML models to be
integrated via EML). Definitely, the cost for data set creation becomes negligible 
over repeated calls of \sfpt.

The time required to found solutions by \emph{FPTuning} varies considerably
depending on the error given as a target (tighter bounds require longer times),
hence the relative differences reported in Tab.~\ref{tab:res_cmp} are more
effective for comparing the approaches. However, it could be useful to provide
some actual numbers to give the order of magnitude. For each benchmark and
computed as average among all error targets, the solution time (in seconds)
required by \emph{FPTuning} are the following: \emph{FWT} $24.3$, \emph{saxpy}
$38.5$, \emph{convolution} $81.9$, \emph{correlation} $180.9$, \emph{dwt}
$81.8$, \emph{BlackScholes} $1512.3$, \emph{Jacobi} $3409.6$.


\begin{table*}
\scriptsize\centering
\begin{tabular}{lcccccccr}
 \toprule
    & \qquad \emph{FWT} & \emph{saxpy} & \emph{convolution} &\emph{dwt} &
    \emph{correlation} & \emph{BlackScholes} & \emph{Jacobi} &
    \qquad \emph{Avg.} \\
    SFPT~vs FPT -- N. Bits (\%) & \qquad 14.1 & -1.0 & 3.7 & 22.6
                   & 14.7 & 22.3 & 29.8  & \qquad 15.2 \\
    SFPT~vs FPT -- Time (\%) & \qquad -62.4 & -14.0 & -33.5 & 53.7 
                              & -79.5 & -80.2 & -65.1 & \qquad \textbf{-55.4} \\
    \midrule
    SFPT+~vs FPT -- N. Bits (\%) & \qquad 4.7 & -3.9 & 0.1 & 0.7
                                     & -1.0 & -1.7 & -4.7  & \qquad
                                     \textbf{-3.0} \\
    SFPT+~vs FPT -- Time (\%) & \qquad -8.2 & 9.1 & 9.1 & -5.8 
                                  & -8.3 & -22.6 & -18.9 & \qquad \textbf{-6.51}
                                  \\
 \bottomrule
\end{tabular}
\caption{Comparison between \sfpt (SFPT in the table), \sfptplus (SFPT+) ~and FPTuning (FPT)}
\label{tab:res_cmp}	
\end{table*}

Concerning the solution quality, \emph{FPTuning} outperforms us (except for
\emph{saxpy}), since our solutions have a higher number of bits (15\% on average).
Conversely, \sfpt~is markedly quicker, as attested by the
average decrease in solution-time of around 55\%. 

\subsection{Extended Approach: \sfptplus} 
\label{sec:exp_staticFPTuner_FPTuning}
As noted in the previous section, \sfpt~is extremely fast but produces
low-quality solutions, as, generally speaking, higher numbers of bits lead to
greater energy consumption. We decided then to extend our approach by combining
our method with \emph{FPTuning}. \emph{FPTuning} algorithm can be decomposed into
two phases: (i) a search for an initial solution satisfying the error
target and (ii) a refinement that iteratively improves the solution
(by lowering the precision through a heuristic), until two
consecutive solutions have the same total number of bits. We propose an extended
approach \sfptplus~that exploits \emph{FPTuning}'s refinement phase
\footnote{https://github.com/minhhn2910/fpPrecisionTuning} to improve the
initial solution found by \sfpt. In practice, \sfptplus~starts from the initial
solution quickly found by \sfpt~and then improves it by attempting to decrease 
the precision of the variables with the heuristic algorithm introduced by \emph{FPTuning}
(a variant of binary search).

The final two rows of Table~\ref{tab:res_cmp} show the results.\\
The time needed by \sfptplus~to find a solution contains an additional component
w.r.t. to \sfpt, namely the time required to improve the initial solution.
\sfptplus~remains faster than \emph{FPTuning} (although the gap is reduced,
average speedup of around 6.5\%) for all but two benchmarks, which are the ones 
with low number of variables (\emph{saxpy} and \emph{convolution}). 
These ``easier'' benchmarks can
be quickly run multiple times; thus, the \emph{FPTuning} approach is less penalizing
-- with applications with more variables \sfptplus~is still significantly
faster, an encouraging sign for the extension of our approach to more complex
programs. More importantly, \sfptplus~also outperforms the SoA in terms of
solution quality; the improvement is relatively small (3\%), but this is
remarkable nonetheless, as experiments performing an exhaustive search on small
benchmarks reveal that \emph{FPTuning} finds solutions very close to the optimal
ones.

\subsubsection{Transfer Learning} \label{sec:exp_transfer_learning}
As mentioned before, at the moment we are mainly interested in a preliminary
evaluation and the comparison
with the SoA, hence we considered a single input set for all 
previous experiments. But at the same time, we want to hint at an additional
benefit that can be obtained with the optimization model w.r.t. \emph{FPTuning}.
Our ML models can learn some of the latent proprieties that characterize the
benchmarks (their precision-error function); some of these relationships may
hold for different input sets. On the contrary, \emph{FPTuning} focuses
exclusively on the problem at hand. Hence, the solutions found by our approach
can be more ``robust'' for different input sets w.r.t. the \emph{FPTuning}
solutions. In a sense, we want to understand if the solution found for a given
input set is transferable to different ones.

We tested this hypothesis in this fashion: I) we generated 30 different input sets
for each benchmark; II) we found the best configurations for a fixed input set
$S_i$ using both \sfpt~and \emph{FPTuning} and for a given error target; III)
finally, we run the benchmark with the configuration just found but feeding it
with the remaining input sets (hence 29 separate runs), and we checked if the
configuration satisfies the error target also for other input sets. For these 
experiments, we considered \sfpt~rather than \sfptplus~since the focus is on 
the solution found by the combination of MP and ML models, without the
added ``noise'' introduced by the heuristic refinement phase of \sfptplus~(the
\emph{FPUtning}-inspired improvement over the first solution found by \sfpt).
The different input sets are vectors of randomly generated numbers. The
solutions for our approach were obtained using data sets of training size equal
to one thousand. Table~\ref{tab:transfer_learn_res} reports the results. Each
row corresponds to an error target; the final one is the average among all
targets.  For each benchmark, we compute the percentage of input sets that
presented an error lower than the target with the configuration found with $S_i$
(excluded from this computation); lower values are preferable since they imply
that the configuration for $S_i$ is more robust. \emph{Blacksholes} and \emph{Jacobi} 
are not reported for space limitations. Columns \emph{FPT} and
\emph{Opt}~(two for each benchmark) indicate, respectively, the results with
\emph{FPTuning} and with \sfpt. The last two columns report the
average values computed among all benchmarks.

\begin{table*}
\scriptsize\centering
\begin{tabular}{lccccccccccrr}
 \toprule
 \multirow{2}{*}{\emph{Target}} & \multicolumn{2}{c}{FWT} 
                                & \multicolumn{2}{c}{saxpy} 
                                & \multicolumn{2}{c}{convolution} 
                                & \multicolumn{2}{c}{dwt} 
                                & \multicolumn{2}{c}{correlation} 
                                & \multicolumn{2}{r}{\emph{All Benchmarks}} \\
             & \quad FPT & \quad Opt 
             & \quad FPT & \quad Opt 
             & \quad FPT & \quad Opt 
             & \quad FPT & \quad Opt 
             & \quad FPT & \quad Opt 
             & \qquad FPT & \quad Opt \\

 0.1 & \quad 10.3 & \quad 44.8 
     & \quad 0 & \quad 0 
     & \quad 17.2 & \quad 0 
     & \quad 62.1 & \quad 79.3 
     & \quad 68.9 & \quad 10.3 
     & \qquad 31.7 & \quad \textbf{26.9} \\
 $10^{-2}$ & \quad 17.2 & \quad 89.6 
           & \quad 0 & \quad 0 
           & \quad 72.4 & \quad 0 
           & \quad 65.5 & \quad 62.1 
           & \quad 68.9 & \quad 13.8 
           & \qquad 44.8 & \quad \textbf{33.1} \\
 $10^{-3}$ & \quad 41.4 & \quad 41.4 
           & \quad 0 & \quad 0 
           & \quad 0 & \quad 0 
           & \quad 65.5 & \quad 86.2 
           & \quad 68.9 & \quad 10.3 
           & \qquad 35.2 & \quad \textbf{27.6} \\
 $10^{-5}$ & \quad 0 & \quad 3.4 
           & \quad 0 & \quad 0 
           & \quad 6.8 & \quad 24.1 
           & \quad 75.9 & \quad 51.7 
           & \quad 79.3 & \quad 62.1 
           & \qquad 32.4 & \quad \textbf{28.3} \\
 $10^{-7}$ & \quad 0 & \quad 65.5 
           & \quad 62.1 & \quad 0 
           & \quad 0 & \quad 17.2 
           & \quad 55.2 & \quad 37.9 
           & \quad 10.3 & \quad 24.1 
           & \qquad \textbf{25.5} & \quad 28.9 \\
 $10^{-10}$ & \quad 0 & \quad 0
           & \quad 0 & \quad 0 
           & \quad 0 & \quad 0
           & \quad 86.2 & \quad 62.1
           & \quad 0 & \quad 20.7
           & \qquad 17.2 & \quad \textbf{16.6} \\
 $10^{-12}$ & \quad 0 & \quad 0
           & \quad 0 & \quad 0 
           & \quad 0 & \quad 27.6
           & \quad 62.1 & \quad 3.4
           & \quad 3.4 & \quad 0
           & \qquad 13.1 & \quad \textbf{6.2} \\
 $10^{-15}$ & \quad 0 & \quad 0
           & \quad 0 & \quad 0 
           & \quad 0 & \quad 0
           & \quad 82.7 & \quad 44.8
           & \quad 0 & \quad 17.2
           & \qquad 16.5 & \quad \textbf{12.4} \\
 $10^{-20}$ & \quad 0 & \quad 0
           & \quad 86.2 & \quad 0 
           & \quad 0 & \quad 0
           & \quad 96.5 & \quad 68.9
           & \quad 0 & \quad 0
           & \qquad 36.5 & \quad \textbf{13.8} \\
 $10^{-25}$ & \quad 0 & \quad 0
           & \quad 6.9 & \quad 0 
           & \quad 0 & \quad 0
           & \quad 7.7 & \quad 0
           & \quad 24.1 & \quad 0
           & \qquad 7.8 & \quad \textbf{0} \\
 \midrule
\emph{Average} & \quad 6.9 & \quad 24.5
           & \quad 15.5 & \quad 0 
           & \quad 9.6 & \quad 6.9
           & \quad 72.4 & \quad 55.2
           & \quad 32.4 & \quad 15.9
           & \qquad 26.1 & \quad \textbf{19.4} \\
  \bottomrule
\end{tabular}
\caption{Transfer Learning Results}
\label{tab:transfer_learn_res}	
\end{table*}

From the table we can see that the ``transferability'' of the solutions strongly
depends on the particular benchmark; for example, \emph{convolution} solutions
are very robust to different input sets, while the contrary happens for
\emph{dwt}. For all benchmarks except \emph{FWT}, \sfpt~is more robust
compared to \emph{FPTuning}; this holds true also if we consider all error
targets (bold values in the last two columns highlight the method with the more
robust solution for a given target). These observations suggest that our
approach is capable of learning part of the underlying patterns that
characterize an application and thus can obtain solutions that can be reused on
different input sets.

However, we are aware that the case of different input sets should be explored
in more detail -- this is a preliminary approach that we plan to improve in
future works. For example, this issue could be dealt with by training the ML
models on multiple samples, representative of the target application; the ML
model may optionally output a probability distribution rather than a single
prediction.

\section{Experimental Results: Energy Efficiency} \label{sec:exp_deploy}

\subsection{Deployment \& Setup} \label{sec:exp_pulp}
Our target platform is
PULPissimo\footnote{https://github.com/pulp-platform/pulpissimo}, an
open-source 32-bit microcontroller based on the RISC-V instruction set
architecture (ISA). This platform supports the R32IMFC ISA configuration,
featuring extensions for integer multiplication and division (``M''),
single-precision FP arithmetic (``F'') and compressed encoding (``C'').  The
core also includes a \emph{smallFloat unit} (SFU)\cite{mach2018}, which provides
a set of non-standard ISA extensions to enable operations on smaller-than-32-bit
FP formats. This unit supports two IEEE standard formats, single-precision
(\emph{binary32}) and half-precision (\emph{16 bit}) ones, and two additional
formats, namely \emph{binary8} and \emph{binary16alt}. The first is an 8-bit
format with low precision (3-bit mantissa), and the second is an alternative
16-bit format featuring a higher dynamic range (8-bit exponent). The SFU also
supports a vectorial ISA extension which makes use of SIMD sub-word parallelism
by packing multiple smaller-than-32-bit elements into a single register; this is
a key feature to reduce energy consumption since it allows to optimize the
circuitry of the HW unit and reduce the memory bandwidth required to transfer
data between memory and registers.

The software ecosystem\footnote{https://github.com/pulp-platform/pulp-sdk}
of the PULP project includes a virtual platform and a compiler (based on GCC
7.1). The virtual platform is cycle-accurate and provides detailed execution
statistics, including instruction and cycle counters, used to evaluate the
energy consumption of the benchmarks. The power numbers have been obtained
through simulation of the post-layout design set to 350 MHz using worst-case
conditions (1.08 V, 125$^{\circ}$C), as detailed in \cite{mach2018}. Finally,
the compiler provides an extended C/C++ type system to make use of the
smallFloat types using additional keywords (\emph{float8}, \emph{float16} and
\emph{float16alt}). The GCC auto-vectorizer has been extended to enable the
adoption of the vectorial ISA extension; whenever reduced-precision variables
can be used, our benchmarks take great advantage of this feature. 

\subsection{Experimental Evaluation} \label{sec:exp_eval}

The energy savings are measured as the energy obtained by running a benchmark
with all single-precision variables (the baseline) over the energy obtained with
the mixed-precision configuration found by \sfptplus; values higher than $1$
indicate energy \emph{gains}, as the mixed-precision approach leads to lower
energy consumption than the baseline. Table~\ref{tab:energy_gain} reports the
results. Each line corresponds to an error bound, and the last line summarizes
the average on all targets; each column reports the energy gain compared to the
baseline.

\begin{table*}
\scriptsize\centering
\begin{tabular}{lcccccccr}
 \toprule
    Error Target & \qquad \emph{FWT} & \emph{saxpy} & \emph{convolution} &\emph{dwt} &
    \emph{correlation} & \emph{BlackScholes} & \emph{Jacobi} &
    \qquad \emph{Avg. over  all benchmarks} \\
    $10^{-1}$ & \qquad 1.00 & 3.99 & 1.35 & 1.00
                        & 1.08 & 1.54. & 2.90 & \qquad 1.84 \\
    $10^{-2}$ & \qquad 1.00 & 2.26 & 1.35. & 1.00
                        & 1.00 & 1.52 & 2.90 & \qquad 1.58 \\
    $10^{-3}$ & \qquad 1.00 & 2.00 & 1.27 & 1.00
                        & 1.00 & 1.29 & 1.74 & \qquad 1.33 \\
    $10^{-4}$ & \qquad 1.00 & 1.90 & 1.22 & 1.00
                        & 1.00 & 1.08 & 1.82 & \qquad 1.29 \\
    $10^{-5}$ & \qquad 1.00 & 2.00 & 1.22 & 1.00
                        & 1.00 & 1.06 & 1.77 & \qquad 1.29 \\
    $10^{-6}$ & \qquad 1.00 & 1.13 & 1.30 & 1.00
                        & 1.00 & 1.00 & 1.78 & \qquad 1.17 \\
    $10^{-7}$ & \qquad 1.00 & 1.00 & 1.00 & 1.00
                        & 1.00 & 1.00 & 1.78 & \qquad 1.11 \\
    \midrule
    Avg. & \qquad 1.00 & 2.04 & 1.25 & 1.00 & 1.00 & 1.21
                                     & 2.09 & \qquad 1.37 \\
 \bottomrule
\end{tabular}
\caption{Energy gains measured as energy consumed with single-precision over
energy with \sfptplus} 
\label{tab:energy_gain}	
\end{table*}

Overall, the results are extremely promising: the average energy gain obtained
with \sfptplus~is 1.37 (around 40\%), and in the benchmarks showing energy
savings the compiler was able to apply automatic vectorization to the code
thanks to the precision-reduction enabled by our tool. However, the gains are
not homogeneous, as for some benchmarks there is no energy saving w.r.t. the
baseline (\emph{FWT, dwt, correlation}); in these cases, the discrete precision
levels offered no margin for energy gain -- more fine-grained mixed-precision
levels could improve this situation and will be investigated in future works.
The results clearly show that, as expected, less strict bounds on the
computation accuracy can ensure higher gains since in these cases the variable
precision can be reduced more markedly.

\section{Conclusion} \label{sec:conclusion} 
In this paper we propose a novel approach for solving the problem of tuning the
precision of FP variables in numerical applications. Our method combines ML
models and an MP optimization model, exploiting the Empirical Model Learning
paradigm. The experimental results reveal that the proposed model is
very fast but, generally speaking,
 produces low-quality solutions. Hence we combine our method with
a refinement algorithm from the literature, thus obtaining an approach that
thoroughly outperforms the SoA.

Moreover, we demonstrate the quality of our approach by measuring the energy gains
obtained via static precision tuning on a virtual platform that emulates
precision-tunable HW, revealing energy savings around 40\% with the static
tuning of FP variables.
\\

\subsection*{\textbf{Acknowledgements}} \noindent This work has been partially
supported by European H2020 FET project OPRECOMP (g.a. 732631).

\bibliographystyle{alpha} \bibliography{bib}

\newcommand{\etalchar}[1]{$^{#1}$}
\begin{thebibliography}{HMWA17}

\bibitem[BLM15]{BonfiettiLM15}
Alessio Bonfietti, Michele Lombardi, and Michela Milano.
\newblock Embedding decision trees and random forests in constraint
  programming.
\newblock In {\em Proceedings of {CPAIOR}}, pages 74--90, 2015.

\bibitem[BLMB11]{BartoliniLMB11}
Andrea Bartolini, Michele Lombardi, Michela Milano, and Luca Benini.
\newblock Neuron constraints to model complex real-world problems.
\newblock In {\em Proceedings of {CP}}, pages 115--129, 2011.

\bibitem[BLMB12]{BartoliniLMB12}
Andrea Bartolini, Michele Lombardi, Michela Milano, and Luca Benini.
\newblock Optimization and controlled systems: {A} case study on thermal aware
  workload dispatching.
\newblock In {\em Proceedings {AAAI}}, 2012.

\bibitem[CBB{\etalchar{+}}17]{chiang2017rigorous}
Wei-Fan Chiang, Mark Baranowski, Ian Briggs, Alexey Solovyev, Ganesh
  Gopalakrishnan, and Zvonimir Rakamari{\'c}.
\newblock Rigorous floating-point mixed-precision tuning.
\newblock {\em ACM SIGPLAN Notices}, 52(1):300--315, 2017.

\bibitem[CN18]{costa2018rbfopt}
Alberto Costa and Giacomo Nannicini.
\newblock Rbfopt: an open-source library for black-box optimization with costly
  function evaluations.
\newblock {\em Mathematical Programming Computation}, 10(4):597--629, 2018.

\bibitem[DK17]{Darulova:2017:TCR:3062396.3014426}
Eva Darulova and Viktor Kuncak.
\newblock Towards a compiler for reals.
\newblock {\em ACM Trans. Program. Lang. Syst.}, 39(2):8:1--8:28, March 2017.

\bibitem[GJea16]{graillat2016promise}
Stef Graillat, Fabienne J{\'e}z{\'e}quel, and et~al.
\newblock Promise: floating-point precision tuning with stochastic arithmetic.
\newblock In {\em Proceedings of the 17th International Symposium on Scientific
  Computing, Computer Arithmetics and Verified Numerics (SCAN)}, pages 98--99,
  2016.

\bibitem[GRG18]{guo2018exploiting}
Hui Guo and Cindy Rubio-Gonz{\'a}lez.
\newblock Exploiting community structure for floating-point precision tuning.
\newblock In {\em Proceedings of the 27th ACM SIGSOFT International Symposium
  on Software Testing and Analysis}, pages 333--343. ACM, 2018.

\bibitem[HHLB11]{hutter2011sequential}
Frank Hutter, Holger~H Hoos, and Kevin Leyton-Brown.
\newblock Sequential model-based optimization for general algorithm
  configuration.
\newblock In {\em International Conference on Learning and Intelligent
  Optimization}, pages 507--523. Springer, 2011.

\bibitem[HMWA17]{ho2017efficient}
Nhut-Minh Ho, Elavarasi Manogaran, Weng-Fai Wong, and Asha Anoosheh.
\newblock Efficient floating point precision tuning for approximate computing.
\newblock In {\em Design Automation Conference (ASP-DAC), 2017 22nd Asia and
  South Pacific}, pages 63--68. IEEE, 2017.

\bibitem[KB14]{kingma2014adam}
Diederik~P Kingma and Jimmy Ba.
\newblock Adam: A method for stochastic optimization.
\newblock {\em arXiv preprint arXiv:1412.6980}, 2014.

\bibitem[KMBC14]{klavik2014changing}
Pavel Klav{\'\i}k, A~Cristiano~I Malossi, Costas Bekas, and Alessandro Curioni.
\newblock Changing computing paradigms towards power efficiency.
\newblock {\em Phil. Trans. R. Soc. A}, 372(2018):20130278, 2014.

\bibitem[LG13]{LombardiG13}
Michele Lombardi and Stefano Gualandi.
\newblock A new propagator for two-layer neural networks in empirical model
  learning.
\newblock In {\em Proceedings of {CP}}, pages 448--463, 2013.

\bibitem[LG16]{LombardiG16}
Michele Lombardi and Stefano Gualandi.
\newblock A lagrangian propagator for artificial neural networks in constraint
  programming.
\newblock {\em Constraints}, 21(4):435--462, 2016.

\bibitem[LMB17]{lombardi2017empirical}
Michele Lombardi, Michela Milano, and Andrea Bartolini.
\newblock Empirical decision model learning.
\newblock {\em Artificial Intelligence}, 244:343--367, 2017.

\bibitem[Mit16]{mittal2016survey}
Sparsh Mittal.
\newblock A survey of techniques for approximate computing.
\newblock {\em ACM Computing Surveys (CSUR)}, 48(4):62, 2016.

\bibitem[MRT{\etalchar{+}}18]{mach2018}
S.~{Mach}, D.~{Rossi}, G.~{Tagliavini}, A.~{Marongiu}, and L.~{Benini}.
\newblock {A Transprecision Floating-Point Architecture for Energy-Efficient
  Embedded Computing}.
\newblock In {\em 2018 IEEE International Symposium on Circuits and Systems
  (ISCAS)}, pages 1--5. IEEE, 2018.

\bibitem[MSea18]{malossi2018transprecision}
A~Cristiano~I Malossi, Michael Schaffner, and et~al.
\newblock {The transprecision computing paradigm: Concept, design, and
  applications}.
\newblock In {\em Design, Automation \& Test in Europe Conference \& Exhibition
  (DATE), 2018}, pages 1105--1110. IEEE, 2018.

\bibitem[MTDM17]{moscato2017automatic}
Mariano Moscato, Laura Titolo, Aaron Dutle, and C{\'e}sar~A Munoz.
\newblock Automatic estimation of verified floating-point round-off errors via
  static analysis.
\newblock In {\em International Conference on Computer Safety, Reliability, and
  Security}, pages 213--229. Springer, 2017.

\bibitem[MTea17]{moscato2017precisa}
Mariano Moscato, Laura Titolo, and et~al.
\newblock Automatic estimation of verified floating-point round-off errors via
  static analysis.
\newblock In Stefano Tonetta, Erwin Schoitsch, and Friedemann Bitsch, editors,
  {\em Computer Safety, Reliability, and Security}, pages 213--229, Cham, 2017.
  Springer International Publishing.

\bibitem[opr]{oprecompProject}
Oprecomp - open transprecision computing.
\newblock {http://oprecomp.eu/}.
\newblock Online; accessed 15 May 2019.

\bibitem[RGNea16]{rubio2016floating}
Cindy Rubio-Gonz{\'a}lez, Cuong Nguyen, and et~al.
\newblock Floating-point precision tuning using blame analysis.
\newblock In {\em Proceedings of the 38th International Conference on Software
  Engineering}, pages 1074--1085. ACM, 2016.

\bibitem[Ste87]{stein1987large}
Michael Stein.
\newblock Large sample properties of simulations using latin hypercube
  sampling.
\newblock {\em Technometrics}, 29(2):143--151, 1987.

\bibitem[TMea18]{tagliavini2018transprecision}
Giuseppe Tagliavini, Stefan Mach, and et~al.
\newblock A transprecision floating-point platform for ultra-low power
  computing.
\newblock In {\em Design, Automation \& Test in Europe Conference \& Exhibition
  (DATE), 2018}, pages 1051--1056. IEEE, 2018.

\bibitem[XMK16]{xu2016approximate}
Qiang Xu, Todd Mytkowicz, and Nam~Sung Kim.
\newblock Approximate computing: A survey.
\newblock {\em IEEE Design \& Test}, 33(1):8--22, 2016.

\end{thebibliography}

\end{document}